\documentclass[twocolumn,showpacs,amssymb,aps, prl]{revtex4}

\usepackage[usenames]{color}

\begin{document}

\title{Conformal Superspace: the configuration space of general
relativity}\author{Julian Barbour}\email{julian@platonia.com}
\affiliation{College Farm, South Newington, Banbury, Oxon, OX15 4JG,
UK}
\author{Niall {\'O} Murchadha}\email{niall@ucc.ie}
\affiliation{Physics Department, University College Cork, Cork,
Ireland}\email{niall@ucc.ie}

\date{\today}

\begin{abstract}
It has long been considered that conformal superspace is the natural configuration space for canonical general relativity. However, this was never definitively demonstrated. We have found that the standard conformal method of solving the Einstein constraints has an unexpected extra symmetry. This allows us to complete the project. We show that given a point and a velocity in conformal superspace, the Einstein equations generate a unique curve in conformal superspace.

\end{abstract}



\maketitle

\vspace{1in}

It was realised more than 50 years ago, due to the pioneering work of Dirac \cite{Dirac}, and Arnowitt, Deser and Misner \cite{ADM}, that general relativity could be expressed as a dynamical theory, just like the other standard theories of physics such as particle mechanics or electromagentism. In GR one can specify initial data and then integrate forward in time. The initial data for gravity consists of a spacelike 3-slice, equipped with a Riemannian 3-metric, $g_{ij}$ and a symmetric tensor $K_{ij}$, which is to be the extrinsic curvature of the slice.
Thus it is the time derivative of the 3-metric. More precisely,
\begin{equation}
{\cal{L}}_{\hat{n}}g_{ij} = 2K_{ij},\label{2}
\end{equation}
where $\cal{L}$ is the Lie derivative and $\hat{n}$ is the unit
timelike normal to the 3-slice. This pair $(g_{ij}, K_{ij})$ can be regarded as
the analogue of the position and momentum in mechanics. GR is
like electromagnetism in that the 10 Einstein equations can be split into 6
evolution equations, which propogate the 3-metric, and 4 constraints
on the initial data. The constraints, which prevent
free specification of $g_{ij}$ and $K_{ij}$, are the main focus of this letter.
For an up-to-date and comprehensive account of the constraints, see
\cite{CBY}, especially Chapter VII.  

The constraints are
\begin{equation}
^{(3)}R - K^{ij}K_{ij} + K^2 = 0, \label{C1}
\end{equation}
\begin{equation}
\nabla_jK_i^{j}  - \nabla_{i}K = 0, \label{C2}
\end{equation}
known as the Hamiltonian and momentum constraints respectively;
$^{(3)}R$ is the scalar curvature of $g_{ij}$ and $K = g_{ij}K^{ij}$ is the
trace of the extrinsic curvature. To identify the `true gravitational degrees of
freedom' and the configuration space of general relativity, we need to `factor'
the metric and extrinsic curvature by the constraints.

We use three spaces: \emph{Riem} is the space of Riemannian
3-metrics $g_{ij\,}(x^k),\, (i,j) =1,2,3,$ on some 3-manifold $\cal
M$ (closed without boundary). The six functions of $x^k$ in
$g_{ij}(x^k)$ define the geometry and the coordinates on it. A
3-geometry has three degrees of freedom at each point and infinitely
many cordinatizations. All explicit calculations will be done in Riem.

\emph{Superspace}, the space of 3-geometries on $\cal M$, is the
quotient of Riem by 3-diffeomorphisms, which eliminate redundant
coordinate information. In its turn, a 3-geometry contains
information of two kinds: the conformal 3-geometry (two degrees of
freedom), which describes angle measurements, and the local scale
factor.

\emph{Conformal Superspace} (CS) is the quotient of superspace by
3-dimensional conformal transformations:
\begin{equation}
\bar{g}_{\,i\,j\,}\,=\,\phi^{\,4}\,g_{\,i\,j\,},~\phi>0.\label{1}
\end{equation}
The fourth power of $\phi$ is just for convenience: the
transformed
3-scalar curvature $R$ is then $\bar{R} = \phi^{-4}R -
8\phi^{-5}\nabla^2\phi$ without additional terms quadratic in
$\nabla\phi$; the preferred power of $\phi$ is two in four
dimensions. 

The idea that conformal superspace was the natural configuration space for
gravity goes back to Lichnerowicz \cite{L}, who showed that using a conformal
transformation one could write the Hamiltonian constraint Eq.(\ref{C1}) as a
nice elliptic equation for the conformal factor. Unfortunately, this
identification was never made precise. We will do so here. This article should
be viewed as a companion to and extension of the seminal work on the subject by
James York \cite{JWY}.

A conformal 3-geometry $\cal C$ is an equivalence class of
3-geometries. It includes infinitely many distributions of the local
scale, which can be seen as coded in $\sqrt g$, where $g$ is the determinant of $g_{ij}$. We wish to treat
$\sqrt g$ as gauge, just like the coordinates. The physical object is
$\cal C$: it is a point in CS, or \emph{shape} of the universe. If we take the trace of Eq.(\ref{2}) we can show
\begin{equation}
{\cal{L}}_{\hat{n}}\sqrt{g} = K\sqrt{g}.\label{H}
\end{equation}
Therefore $(\sqrt{g}, K)$ are canonically conjugate variables. 
If we think of the volume element, $\sqrt{g}$, as gauge, we must think of $K$ as one also.

It turns out that \emph{transverse-traceless} (TT) tensors (a tensor $h^{TT}_{ij}$ is TT if it is both \emph{transverse}
($\nabla^jh^{TT}_{ij}=0$) and \emph{traceless}
($g^{ij}h^{TT}_{ij}=0$)) are natural objects on CS. 

First, TT tensors define the tangent space to CS. Consider two nearby metrics 
in Riem, $g_{ij}$ and $g_{ij} + \epsilon h_{ij}$, with $\epsilon$ a small
parameter. Any symmetric tensor, in particular $h_{ij}$,  has a unique
TT part  with respect to a $g_{ij}$ via \cite{LY}
\begin{equation}
h_{ij} = h^{TT}_{ij} + \nabla_{i}\lambda_j  + \nabla_{j}\lambda_{i} + {1 \over
3}hg_{ij};\label{karel}
\end{equation}
$\nabla_{i}\lambda_j +
\nabla_{j}\lambda_{i}$ is the change in $g_{ij}$ due to an
infinitesimal change of coordinates, and the final (scalar) term is
an infinitesimal conformal transformation. These two are therefore
the generators of the symmetries we factor out in going from Riem to CS, and merely change the representation of the $\cal C$ corresponding to the given
$g_{ij}$. The perturbation of $\cal C$ is $h^{TT}_{ij}$. Because its magnitude can be changed by multiplying by a constant
$c$ without changing its essential nature (if
$h^{TT}_{ij}$ is $TT$, so is $ch^{TT}_{ij}$), it is a
tangent vector.

Second, $TT$ tensors are \emph{conformally covariant} \cite{LY}. If $h^{TT}_{ij}$
is $TT$ wrt $g_{ij}$ and $\xi$ is any function, then
$\xi^{-2}h^{TT}_{ij}$ is $TT$ with respect to $\xi^4g_{ij}$. This is just straightforward algebra. The exponent $-2$ is the power of the conformal factor for a down-down TT tensor, it is $-6$ for up-down and $-10$ for up-up. Thus
$(\xi^4g_{ij},\xi^{-2}h^{TT}_{ij})$ represents the same $\cal C$ and
tangent vector as $g_{ij}$ and $h^{TT}_{ij}$. 

Third, as first pointed out by Peter Higgs \cite{Higgs}, if we have an action on
superspace, the associated momentum must be divergence-free. If the action is
defined on conformal superspace, the momentum must also be tracefree. Indeed,
the variation of the action with respect to a change in the metric can be
written
$$
\delta L = \int \pi^{ij} \delta g_{ij} d^3x,
$$ 
where $\pi^{ij}$ is the momentum conjugate to $g_{ij}$. If the change of the
metric is just a conformal change, i.e., $\delta g_{ij} = (\delta f) g_{ij}$,
and if the action is invariant, we get
$$
\delta L = 0 = \int \pi^{ij} (\delta f) g_{ij} d^3x.
$$
If this vanishes for all $\delta f$, we must have $\pi^{ij}g_{ij} = 0$.

Fourth, asymptotically flat initial data have a well-defined total energy, the
ADM energy. Brill and Deser, in \cite{BD}, showed that if one made a
perturbation expansion in the initial data about flat space, the first
non-trivial contribution to the energy came at second order and had the form
$$
16\pi({1 \over 2} \delta^2E_{ADM}) = \int d^3x[{1 \over 4}(\delta
g^{TT}_{ij,k})^2 + (\delta K^{TT}_{ij})^2].
$$
This expression is very similar to the Poynting energy expression, $(E^2 + B^2)$, in electromagnetism, and shows that the $TT$ terms are the true excitations of the gravitational field in the weak-field limit. The challenge is to extend this to the strong field/no boundary case.

The standard conformal method, \cite{CBY}, starts with the realization that if the extrinsic curvature has a constant trace then the momentum constraint, Eq.(\ref{C2}), implies that the tracefree part is transverse. This means that we can write
\begin{equation}
K_{ij} = K_{ij}^{TT} + {1\over 3}Kg_{ij}.\label{3}
\end{equation}
We start with freely specifiable $(\bar g_{ij},\bar K_{ij})$ and make the
decomposition (\ref{karel}), obtaining $(\bar g_{ij},\bar K_{ij}^{TT})$. We call
this pair \emph{the initial data}, conceived as a point and tangent vector in
CS but of necessity represented as a point and tangent vector in Riem. We
adjoin a constant $K$, obtaining the triplet $(\bar g_{ij}\,,\bar
K^{TT}_{ij}, K)$. We
now seek a conformal factor $\phi$ which maps this triplet into a new triplet
satisfying the constraints via  $( g_{ij}\,,
K^{TT}_{ij}, K) = (\phi^4\bar g_{ij}\,,\phi^{-2}\bar K^{TT}_{ij}, K)$.
Then $K_{ij}$, constructed  as $K_{ij}= K_{ij}^{TT} + {1\over 3}Kg_{ij}$,
following from Eq.(\ref{3}), satisfies the momentum constraint, while the
Hamiltonian constraint transforms into the Lichnerowicz - York (L-Y) equation
\begin{equation}
8{\bar\nabla}^2\phi - {\bar R}\phi +{\bar K}^{ij}_{TT}{\bar K}^{TT}_{ij}\phi^{-7} - {2 \over 3}{K}^2\phi^5 = 0; \label{LYE}
\end{equation}
$\bar R$ is the 3-scalar curvature formed from $\bar g_{ij}$ (as is
$\bar\nabla$). Equation (\ref{LYE}) always has a unique positive solution $\phi>0$
\cite{NY} as long as $K \neq 0$ and $\bar{K}^{TT} \neq 0$.  

Further, we  can transform the initial data with an arbitrary positive function
$\xi$ to $(\bar g'_{ij}\,,\bar K^{'TT}_{ij}, K')=(\xi^4\bar
g_{ij}\,,\xi^{-2}\bar K^{TT}_{ij}, K)$. The conformal covariance of the L-Y
equation emerges via the fact that when these `new' data are injected  into the
L-Y equation the `new' conformal factor $\phi' = \phi/\xi$! This means that the
data we construct to satisfy the constraints $( g'_{ij}\,, K^{'TT}_{ij},
K')=(\phi'^4\bar g'_{ij}\,,\phi'^{-2}\bar K^{'TT}_{ij}, K')$ are identical to
the set we got without the transformation with $\xi$. This is {\it almost} good
enough: since making an arbitrary
conformal transformation changes nothing, $\bar g_{ij}$ can be regarded as a
point in CS and $\bar K_{ij}^{TT}$ can be regarded as a velocity in CS at that
point. However, the need to specify $K$ as an extra initial datum complicates
things. In fact, this is unnecessary because the constraints turn out
to have an extra, unexpected but simple, symmetry. 

Pick a (positive or negative) constant $A$. Let
$(g_{ij}, K_{ij})$ solve the constraints. Now transform them as follows:
$(\bar{g}_{ij}, \bar{K}_{ij}) = (A^2g_{ij}, AK_{ij})$. The new data will also
satisfy the constraints. Each term in the Hamiltonian constraint picks up a
factor of $A^{-2}$ and each term in the momentum constraint is multiplied by
$A^{-1}$.

This symmetry also commutes with the conformal method of constructing solutions
to the constraints as follows: Let us take the specified initial data and
transform them as follows: pick a (positive or negative) constant $A \neq 0$.
Construct `new' initial data (we think of these as `rescaled' data, the
terminology will become clear soon) $(\bar g'_{ij}\,,\bar K'^{TT}_{ij},
K')=(A^2\bar g_{ij}\,,A\bar K^{TT}_{ij}, K/A)$. Substitute these data into the
L-Y equation. One can see that each term in the equation picks up a factor of
$A^{-2}$. Therefore $\phi' = \phi$. Hence this rescaling commutes
with the L-Y equation. We can rescale  either before or after solving the L-Y
equation. We get the same final (rescaled) data satisfying the constraints.

Why `rescaling'? Take any initial data satisfying the constraints and propagate
them. This gives a (patch of) spacetime with a spacetime 4-metric $g_{\mu\nu}$
satisfying the Einstein equations. If we use geometric units, so that the speed
of light  = 1, then we have only one dimensionful quantity (say `meters').
Following Dicke \cite{RD} we choose to put the dimensions into the metric and
consider the coordinates as pure numbers, labels of points. Let us decide to
change our units from `meters' to `yards'. This is achieved by multiplying the
spacetime metric by a spacetime constant $A$, i.e., $g_{\mu\nu} \rightarrow
A^2g_{\mu\nu}$. This new metric continues to satisfy the Einstein equations. The
effect of this rescaling on the 3+1 data is $(g'_{ij}, K'_{ij}) = (A^2g_{ij},
AK_{ij})$, or $(g'_{ij}, K^{'TT}_{ij}, K') = (A^2g_{ij}, AK^{TT}_{ij}, K/A)$.

We should stress that this `rescaling' transformation $(g'_{ij}, K^{'TT}_{ij},
K') = (A^2g_{ij}, AK^{TT}_{ij}, K/A)$ is \emph{not} a subset of the conformal
transformations $(\bar g'_{ij}\,,\bar K^{'TT}_{ij}, K')=(\xi^4\bar
g_{ij}\,,\xi^{-2}\bar K^{TT}_{ij}, K)$ mentioned earlier. In one case the
solution of the constraints that emerges is rescaled, in the other case the
solution is unchanged. We now show that this new extra symmetry means that $K$ 
does not correspond to an extra physical initial datum in CS but merely to a
choice of units in spacetime. 

We picked the initial data as a point in Riem (as Riemannian 3-metric
$\hat{g}_{ij}$) and a TT tensor $\hat{K}^{TT}_{ij}$ on $\hat{g}_{ij}$,
regarding these as a point and tangent in CS even though we have to work in
Riem. Now we pick a constant $K_1$, which may be positive or negative, but not
zero.  From these we construct `intermediate' data
$(\bar g_{ij}\,,\bar K^{TT}_{ij}, \bar{K})=(K_1^{-2}\hat g_{ij}\,,K_1^{-1}\hat
K^{TT}_{ij}, K_1)$. These intermediate data are of the standard form, i.e.,
metric + TT tensor + constant, so we can substitute them into the L-Y equation,
Eq.(\ref{LYE}), find the solution $\phi_1$ and construct data which satisfy
the constraints $( g_{ij}\,,K^{TT}_{ij}, K) = (\phi_1^4\bar
g_{ij}\,,\phi_1^{-2}\bar K^{TT}_{ij}, \bar K_1)$. 

Let us now go back and, leaving the initial data unchanged, pick a new constant,
$K_2$, and repeat the construction. We find new intermediate data $(\bar
g_{ij}\,,\bar K^{TT}_{ij}, \bar{K})=(K_2^{-2}\hat g_{ij}\,,K_2^{-1}\hat
K^{TT}_{ij}, K_2)$, a new solution $\phi_2$ to the L-Y equation, and new
solution data satisfying the constraints. What is the relationship between the
two sets of solution data? If we look at the two sets of intermediate data we
can see that the mapping between them is a rescaling transformation as
introduced earlier. More precisely, $(K_2^{-2}\hat g_{ij}\,,K_2^{-1}\hat
K^{TT}_{ij}, K_2) =(A^2K_1^{-2}\hat g_{ij}\,,AK_1^{-1}\hat K^{TT}_{ij}, K_1/A)$
with $A = K_1/K_2$. This means that $\phi_1 = \phi_2$, and, one of the solutions
of the constraints is just a rescaling of the other. Therefore, holding the
initial data fixed, and changing the value of $K$ generates solutions of the
constraints that are related by rescaling.

We can see three routes to proceed from this point. The first, and for us least
desirable, is to abandon the 3 + 1 viewpoint and return
to a 4-dimensional picture. Then each set of initial data will generate
a spacetime satisfying the vacuum Einstein equations, and each
spacetime generated from a given set of initial data can be rescaled as above
into
another that satisfies the same equations.

The second is to maintain the 3 + 1 idea, but live with many-fingered time. From a given set of initial data, we know that the evolution equations generate an infinite family of curves through superspace, each corresponding to a different slicing of the spacetime. The family of curves arising from data set 1 is different from the curves from data set 2. However, when the families are mapped into conformal superspace, they coincide.

The third, and the one we favour, is to realise that we have constructed a CMC initial data slice, and that it is very natural to extend this into the spacetime as a CMC foliation. Look at Eq.(\ref{H}), 
${\cal{L}}_{\hat{n}}\sqrt{g} = K\sqrt{g}$.
This tells us that, on a CMC slice, the fractional time rate of change of the
local volume is a constant. Therefore these CMC slices are the natural `Hubble
time' slices of a cosmology. There always exists a (two-sided) CMC foliation
around any given CMC slice. This, and only this, preserves the TT-ness of the
extrinsic curvature. To maintain it, we solve the elliptic lapse-fixing equation
\begin{equation}
\nabla^2N - K^{ij}K_{ij} N =C\label{lfe}
\end{equation}
for the function $N$, the lapse function; $C$ is some constant, conveniently taken to be
$C=-1$. Equation (\ref{lfe}) has a unique solution. In addition, if $C<0$, then
$N>0$ and vice versa.

We now evolve $(g_{ij}, K_{ij})$ with respect to the time label $t$ using
\begin{equation}
{\partial g_{ij} \over \partial t} = 2NK_{ij} +\nabla_{i} N_j + \nabla_{j} N_{i}.\label{NK}
\end{equation}
This is just rewriting Eq.(\ref{2}) in 3 + 1 language. There is also
 a further equation for $\partial K_{ij}/\partial t$ that we can
omit; in both we may set the freely specifiable $N_j$ to zero, but
must continuously update $N(t,x^k)$ using Eq. (\ref{lfe}). This evolution system fixes $\partial K/\partial t = C$ and generates the desired CMC foliation. 

Thus each choice of initial data with arbitrary, but constant, $K$, will
generate different (rescaled) paths through superspace but will generate the
\emph{same} path through conformal superspace.

Let us point out that no natural `arrow of time' emerges in this analysis. The
key point is that the rescaling constant $A$ can be either positive or negative.
The metric remains Riemannian, with signature (+, + ,+), because it is
multiplied by $A^2$, but the extrinsic curvature changes sign. This
merely exchanges what we think of as `future' with `past'. 

We choose the initial data to be a point and a velocity in conformal superspace,
with no specification of the local scale. This local scale emerges when we solve
the L-Y equation. To find initial data in Riem or in superspace we also
need to specify a `unit length'. This is why we have to pick a $K$.  However,
the solution to the Einstein equations, regarded as a curve  in conformal
superspace, is independent of the choice of $K$.

We define \emph{shape dynamics} as a theory in which a point ${\cal
C}_0$ in CS and any one of ${\cal C}_0$'s tangent vectors together determine
a unique curve in CS. The initial data are thus a conformal
equivalence class $(\xi^4g_{ij}\,,\xi^{-2}K^{TT}_{ij})$. We have shown that GR
is such a theory. We emphasize that the analogue of (\ref{C1}) in
\cite{CS}, in which only \emph{directions} in CS are dynamic, leads to
an emergent local scale without `expansion of the universe'. When $K$ is in the
algorithm, this allows tangent vectors to be dynamic, and global expansion
emerges too. In \cite{CS+V} we construct an action which naturally generates G
R in the CMC gauge.

We see a strong case for regarding local shape as the core of GR, and local
scale as mere gauge. The conceptual basis of shape
dynamics is minimal, matches what can actually be observed,
and allows all the unquestionably physical solutions
of Einstein's full theory while excluding many (such as those with
closed timelike curves) that seem clearly unphysical. In fact, we
considered only vacuum solutions here, but matter, both fundamental
and phenomenological, can be included in York's method, as can open
universes by specifying spatial boundary conditions \cite{IMY}. At
the same time, our mathematical framework is fully adequate to
describe all cosmological observations, e.g., red shift
determinations are \emph{local} comparisons of galactic and
laboratory wavelengths. Finally, we believe that the identification of shape
dynamics as the core of GR
removes the motivation for the more complicated conformal theories
advocated in \cite{Weyl}. One of us has recently
argued that the result developed here implements
Mach's principle in general relativity \cite{JB}

We think shape dynamics could have consequences in cosmology but
that it would be premature to try to identify them here. In
contrast, there are clear possibilities in the quantum domain. The
time at which wave-function collapse occurs is ill defined in
relativity; however the simultaneity associated with the shape-dynamic CMC
foliation could bring
interpretational clarity. This is also true for the `problem of
time' \cite{POT}, which arises from the
ambiguity in the time evolution in superspace if foliation
invariance (many-fingered time) is made inviolate. Moreover,
insistence on this has hitherto made it impossible to identify the
space on which the putative wave function of the universe should be
defined. In shape dynamics, the space must be
conformal superspace.

\begin{acknowledgments}

We thank Jonathan Halliwell for an email exchange that prompted this analysis; JB thanks Edward Anderson, Tim Clifton, Pedro
Ferreira, Henrique Gomes, Sean Gryb, Karel Kucha\v r, and Erhard Scholz for
discussions. Boris Barbour and D.J. Thornton have made many helpful comments on our text. JBB
is supported by Grant RFP2-08-05 of the Foundational Questions
Institute (fqxi.org) and N\'OM by Grant 07/RFP/PHYF148 from Science
Foundation Ireland.

\end{acknowledgments}

\end{document}